# Revealing nonvolatile behaviors in magneto-thermal switching using microstructure-controlled superconducting composites

Keigo Ito,[1,2] Yui Sakamoto,[2] Hossein Sepehri-Amin,[2] Yuto Watanabe,[3] Poonam Rani,[3] Kumpei Imamura,[1,4] Takamasa Hirai,[1,2] Keisuke Hirata,[1,2] Shunsuke Mori,[1,2] Yusuke Nakanishi,[1,2] Kenichiro Hashimoto,[1,4] Takasada Shibauchi,[1] Yoshikazu Mizuguchi,[3] Ken-ichi Uchida,[1,2,*] and Fuyuki Ando[2,*]

*[1]Department of Advanced Materials Science, Graduate School of Frontier Sciences, The University of Tokyo, 5-1-5 Kashiwanoha, Kashiwa, Chiba 277-8561, Japan*
*[2]Research Center for Magnetic and Spintronic Materials, National Institute for Materials Science, 1-2-1 Sengen, Tsukuba, Ibaraki 305-0047, Japan*
*[3]Department of Physics, Tokyo Metropolitan University, 1-1 Minami-osawa, Hachioji, Tokyo 192-0397, Japan*
*[4]Department of Physics, Kyoto University, Sakyo-ku, Kyoto, Kyoto 606-8502, Japan*

Thermal conductivity in a conductor changes by the application of an external magnetic field, which functions as a magneto-thermal switch. For superconductors, a large magneto-thermal switching can occur through a superconducting-to-normal conducting phase transition due to the change in the electron contribution in thermal conductivity. Arima et al. recently reported a nonvolatile nature of the magneto-thermal switching for superconducting solders, which consist of phase-separated Sn and Pb domains. Although they clarified that magnetic flux trapping is required to induce the nonvolatile magneto-thermal switching, a rule for such material design is still unclear. Here, we investigate the microstructure dependence of magneto-thermal switching in superconducting Sn/Pb multilayered composites, which are created by an accumulative roll bonding method. The thickness of each layer, that is the scale of microstructure, can be systematically controlled by the repetition number of roll bonding while the whole sample size and average composition are unchanged. We find that, as the formation of micro-scaled Sn domains proceeds by increasing the repetition number, a nonvolatility in the magneto-thermal conductivity gradually appears in correlation with the remanent magnetization. This study directly confirms that the inclusions with a size comparable to or less than the magnetic vortex in superconducting matrix is essential for magnetic flux trapping, enabling the nonvolatile magneto-thermal switching in superconducting composites.

## I. INTRODUCTION

Magneto-thermal switch is a device where thermal conductivity $\kappa$ changes by applying an external magnetic field or changing the magnetization direction [1,2]. Magnetic and superconducting materials have been reported to show unique magneto-thermal switching (MTS) behaviors to date. First, ferromagnetic materials generally induce MTS driven by several mechanisms involving the magneto-thermal resistance effects [3–5] and magnon thermal conduction [6–10]. The anisotropic magnetoresistance effect in ferromagnetic materials varies an electrical resistivity depending on the relative angle between magnetization and charge current directions. MTS is also driven through the modulation of electron thermal conductivity by the anisotropic magneto-thermal resistance effect as the thermal analogue [3]. Also, MTS was observed in an artificial superlattice composed of nanoscale ferromagnetic and non-magnetic layers [4,5]. When the magnetization directions of adjacent ferromagnetic layers are parallel, $\kappa$ increases, while $\kappa$ decreases when antiparallel. Meanwhile, MTS is induced in superconductors when an external magnetic field exceeding the critical field is applied [11–15]. In metallic superconductors, through the superconducting-to-normal conducting phase transition, electrons become to contribute to the thermal conduction, thereby increasing $\kappa$. Magneto-thermal switching ratio (MTSR) is defined as the ratio of thermal conductivity in the high conductivity state ($\kappa_{high}$) to that in the low conductivity state ($\kappa_{low}$), i.e.,

*Contact author: ANDO.Fuyuki@nims.go.jp, UCHIDA.Kenichi@nims.go.jp

$$\mathrm{MTSR} = \frac{\kappa_{\mathrm{high}} - \kappa_{\mathrm{low}}}{\kappa_{\mathrm{low}}}. \quad (1)$$

Among all the materials reported to date, metallic superconductors are known to exhibit the highest MTSR.

Recently, a nonvolatile nature in MTS has been discovered in metallic superconductors [16–18]. This refers to a behavior that, after the superconducting state is destroyed by the application of an external magnetic field **H** whose magnitude $H$ is higher than the critical magnetic field, the $\kappa$ value is maintained higher than $\kappa_{\mathrm{low}}$ even when $H$ is returned to zero. This nonvolatile MTSR is defined as,

$$\text{Nonvolatile MSTR} = \frac{\kappa_{\mathrm{fin}}(H = 0\,\mathrm{Oe}) - \kappa_{\mathrm{low}}(H = 0\,\mathrm{Oe})}{\kappa_{\mathrm{low}}(H = 0\,\mathrm{Oe})}, \quad (2)$$

where $\kappa_{\mathrm{fin}}$ is $\kappa$ after experiencing $H$ larger than the critical magnetic field. Such nonvolatile MTS has been observed in type-II superconductors [15] and phase-separated composites [16,17]. It is anticipated that the generation and pinning of magnetic flux in type-II superconductors, leading to the formation of partial normal conducting region, correlates with nonvolatile MTS. On the other hand, it has been reported that even in phase-separated composites of type-I superconductors Sn and Pb, magnetic flux is trapped and nonvolatile MTS occurs. Specifically, when $H$ exceeding the critical magnetic field of Pb is applied and then removed, magnetic flux is trapped in the Sn region which behaves like a type-II superconductor [19]. Consequently, even at zero magnetic field, the Sn region with magnetic flux coexists with the superconducting Pb region. Thus, the generation of nonvolatile MTS for the Sn-Pb system owes to thermal conduction in the Sn region where superconducting state is broken. However, no systematic experimental investigation was conducted to reveal the requirement for such magnetic flux trapping and guidelines of material engineering for nonvolatile MTS.

We report that the development of microstructure is essential for inducing nonvolatile MTS in superconducting composites. To systematically investigate the composite structure dependence of magneto-thermal conductivity, Sn/Pb multilayered composite samples were created by an accumulative roll bonding (ARB) method, which is a technique for producing bulk metallic composite materials by laminating several kinds of metal plates under compressive pressure using two rollers [20–24]. An advantage of the ARB method is that only the microstructure can be controlled in the wide scale range by varying the repetition number $n$ of roll bonding without changing the whole sample size and average composition [Fig. 1(a)]. An as-bonded Sn/Pb bilayer ($n = 1$) exhibits typical type-I superconducting behavior, that is, no hysteresis in magneto-thermal conductivity. Interestingly, as microstructural refinement progresses with the increase of $n$, a nonvolatile behavior in magneto-thermal conductivity gradually emerges and the nonvolatile MTSR monotonically increases. The remanent magnetization also exhibits a similar trend on $n$. This finding suggests that the formation of microstructure plays an important role in the emergence and amplitude of nonvolatile MTS.

## II. METHODS

In this study, we created Sn/Pb multilayered composites with $n$ = 1, 4, 7, 10, and 13 by the ARB method. 99.9% (3N) purity Sn and Pb plates (Sasaki Solder Industry Co., Ltd.) were used as raw materials. The Sn and Pb plates were initially cut into an equal dimension of 0.5 mm × 35 mm × 35 mm. As shown schematically in Fig. 1(a), the following four processes were repeatedly performed: (1) Pre-annealing the stacked Sn and Pb plates at 150°C for 3 min in a $N_2$ atmosphere to ensure the mechanical bonding preventing from a surface oxidization, (2) roll bonding with the roller temperature of 150°C and roller speed of 1.0 m/min in an ambient temperature and pressure, (3) cutting the bonded sample into two equal halves, and (4) manually stacking the cut samples. Finally, all the samples were rolled to have a uniform thickness of 0.3 mm. Note that, in general, the method to create nonvolatile MTS materials is not limited to the ARB method. As long as phase-separated systems, one can adopt versatile synthesis approaches including melting and sintering methods.

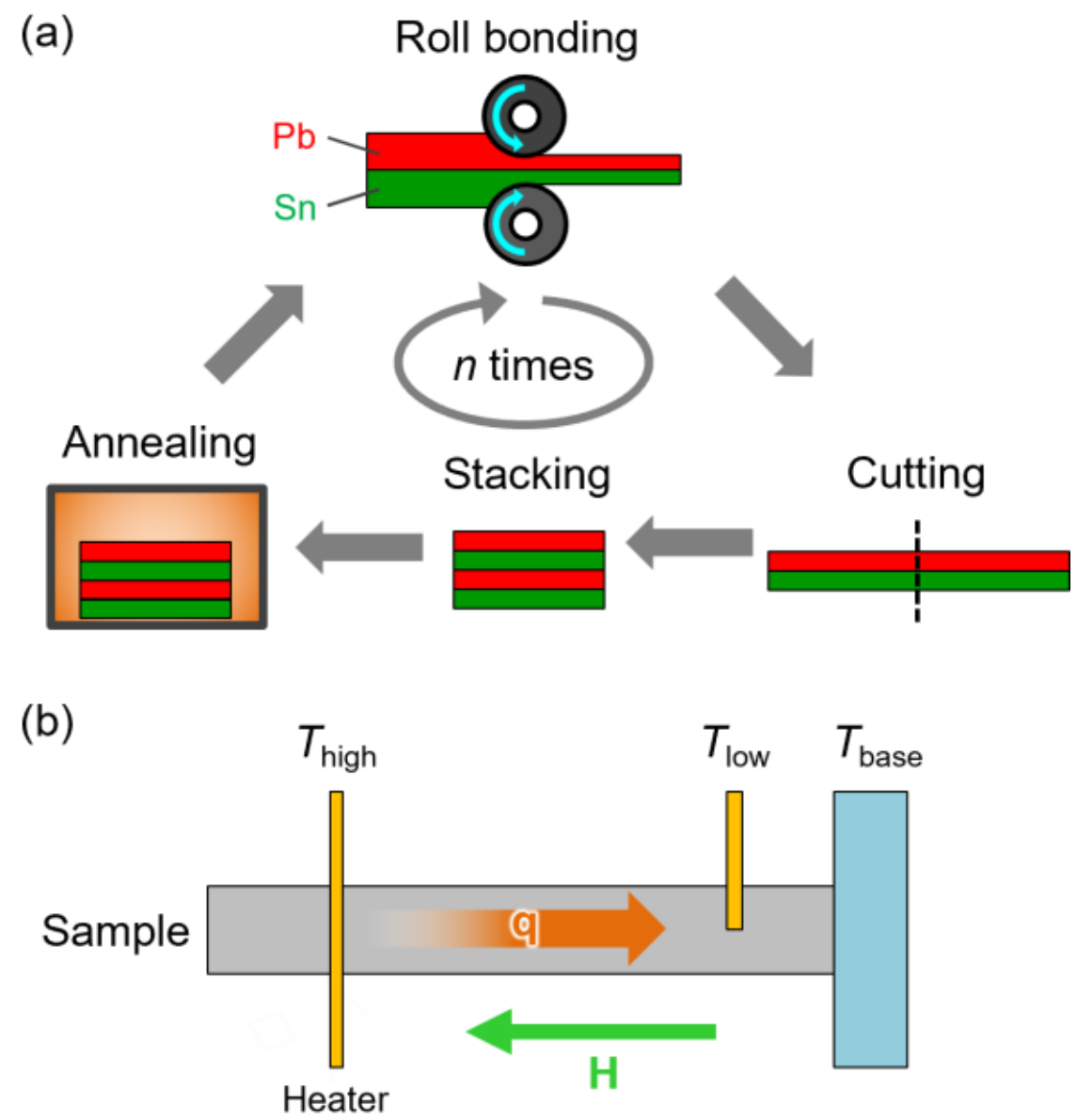

FIG 1. (a) Schematic of the ARB process to create Sn/Pb multilayered composites. The microstructure is systematically controlled by the repetition number $n$. (b) Schematic of the setup for thermal conductivity measurement based on steady-state heat flow method. Hot and cold sides temperatures, $T_{\mathrm{high}}$ and $T_{\mathrm{low}}$, and a base temperature $T_{\mathrm{base}}$ are measured while applying a heat flux **q** by a heater under an external magnetic field **H**.

*Contact author: ANDO.Fuyuki@nims.go.jp, UCHIDA.Kenichi@nims.go.jp

We performed microstructure analysis and thermal conductivity, specific heat, and magnetization measurements for the prepared Sn/Pb multilayered composites. The cross-sectional microstructure and elemental mapping were observed using a scanning electron microscopy (SEM) with an energy-dispersive X-ray spectroscopy (EDX) (Cross-Beam 1540ESB, Carl Zeiss AG). Thermal conductivity measurements were conducted in a Physical Property Measurement System (PPMS, Quantum Design) with a thermal transport option using a four-probe steady-state method with heater, two thermometers, and base-temperature terminal [Fig. 1(b)]. An external magnetic field $\mathbf{H}$ is applied parallel to a heat flux $\mathbf{q}$ with a magnitude $q$. The sample was connected to thermometers using Cu wires and Ag paste to measure the temperature gradient $\nabla T$. Then, $\kappa$ is characterized from $q$ and $\nabla T$ as

$$\kappa = \frac{q}{\nabla T}. \quad (3)$$

Specific heat measurements were performed in PPMS with a heat capacity option using a relaxation mode. Magnetization measurements were performed by a superconducting quantum interference device (SQUID) magnetometry on Magnetic Property Measurement System (MPMS3, Quantum Design) with a VSM mode.

## III. RESULTS

### A. Cross-sectional Microstructure

Figure 2 shows the SEM-EDX elemental maps of Pb and Sn obtained from the cross-section of the Sn/Pb multilayered composites with different $n$. As $n$ increases from 1 in Fig. 2(a) to 13 in Fig. 2(e), the average thickness of Sn and Pb layers monotonically decreases and a partial fragmentation is observed in the samples with $n = 10$ and 13, indicating the progress of the microstructure refinement. These series samples have a significant difference in their composite structure scale, where the average thickness ranges from 150 μm for $n = 1$ [Fig. 2(a)] to less than 1 μm for $n = 13$ [Fig. 2(e)]. Figure 2(f) shows the composition profiles of Pb and Sn for the sample with $n = 13$ [Fig. 2(e)], which indicates that Pb and Sn maintain the ideally phase-separated state without alloying during the ARB process.

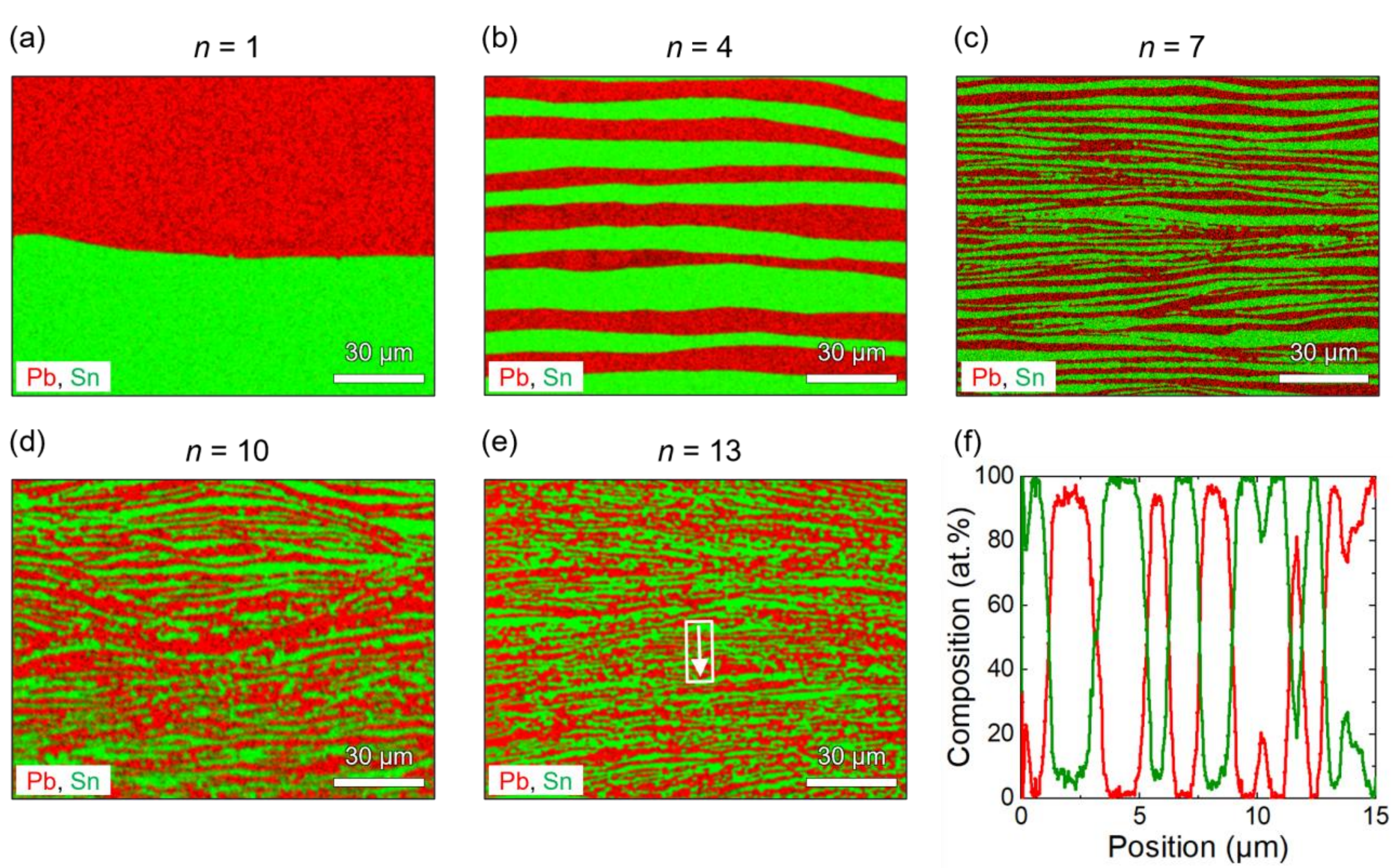

FIG 2. Cross-sectional SEM-EDX elemental maps of Pb and Sn obtained from the Sn/Pb multilayered composites with the repetition number $n$ of (a) 1, (b) 4, (c) 7, (d) 10, and (e) 13. (f) Composition profiles of Pb and Sn in the sample with $n$ = 13 along the white arrow marked in (e).

*Contact author: ANDO.Fuyuki@nims.go.jp, UCHIDA.Kenichi@nims.go.jp

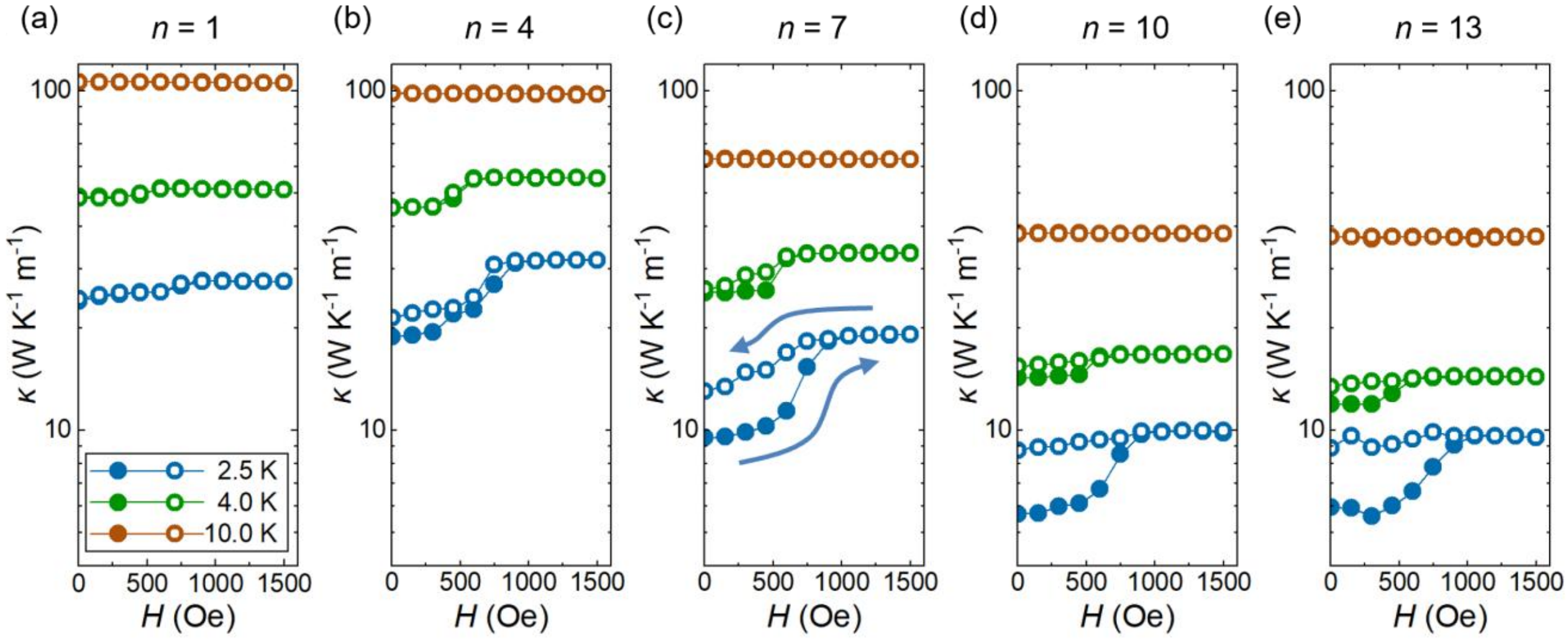

FIG 3. Magnetic field $H$ dependence of the thermal conductivity $\kappa$ at 2.5, 4.0, and 10.0 K for the Sn/Pb multilayered composites with the repetition number $n$ of (a) 1, (b) 4, (c) 7, (d) 10, and (e) 13. The closed circles indicate the data obtained with increasing $H$, whereas the open circles with decreasing $H$. Arrows in (c) also indicates the sweeping direction of $H$.

### B. Thermal Conductivity Measurement

Figure 3 shows the measurement results of $\kappa$ for all the Sn/Pb multilayered composites. The base temperature $T_{\mathrm{base}}$ was set to 10.0 K, 4.0 K, or 2.5 K from 10.0 K under $H = 0$ Oe, then $H$ increased from 0 to 1500 Oe and returned to 0 Oe. During this process, $\kappa$ was measured when $H$ was stabilized at each value. As shown in Figs. 3(a)-(e), $\kappa$ decreases as the temperature decreases from 10.0 to 2.5 K, consistent with the previous reports [16,17]. Given the superconducting transition temperatures of Sn and Pb are 3.7 K and 7.2 K, respectively, both materials are in normal conducting states at 10.0 K, only Pb is superconducting at 4.0 K, and both Sn and Pb are superconducting at 2.5 K. At 10.0 K, $\kappa$ exhibits almost no $H$ dependence for all the samples due to the absence of phase transition. Meanwhile, $\kappa$ at 10.0 K monotonically decreases with increasing $n$ associated with the increased scattering of electrons and phonons at the multiple interfaces [23,24]. On the other hand, at 4.0 K and 2.5 K, the increase of $H$ causes the steep enhancement of $\kappa$, where the superconducting state is destroyed when $H$ exceeds the critical magnetic field of Pb. Interestingly, while the nonvolatile behavior in magneto-thermal conductivity is absent for the Sn/Pb bilayer sample with the Sn and Pb thicknesses of 150 μm in Fig. 3(a), it gradually appears from the result at 2.5 K for the sample with $n = 4$ in Fig. 3(b), which has the average thickness less than 20 μm comparable order of the magnetic vortex in solders [19]. The nonvolatile behavior becomes more pronounced as $n$ increases more than 4. When $n \geq 7$ (the average thickness less than 3 μm), nonvolatile MTS is more clearly observed both at 2.5 K and 4.0 K in Fig. 3(c)-(e).

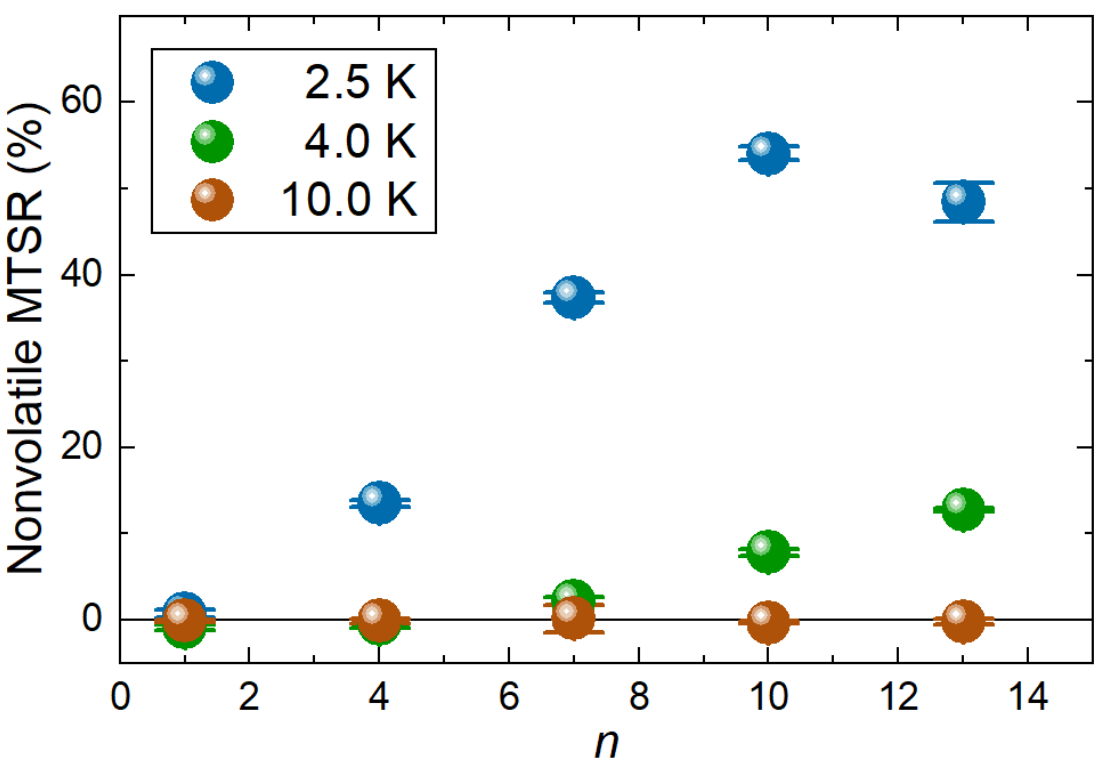

FIG 4. The repetition number $n$ dependence of nonvolatile MTSR for the Sn/Pb multilayered composites at 2.5, 4.0 and 10.0 K.

*Contact author: ANDO.Fuyuki@nims.go.jp, UCHIDA.Kenichi@nims.go.jp

In Fig. 4, we plot nonvolatile MTSR. Nonvolatile MTS is observed only at 2.5 and 4.0 K below the critical temperature of Pb (7.2 K). The nonvolatile MTSR value monotonically increases from 0% for the Sn/Pb bilayer sample ($n$ = 1) to a maximum of approximately 50% at 2.5 K as $n$ increases. At 2.5 K, the nonvolatile MTSR value almost saturates around $n$ = 10 because only a partial fragmentation progressed by ARB without decreasing the domain size for $n \geq 10$. Meanwhile, nonvolatile MTSR at 4.0 K is lower than that at 2.5 K because of the thermal instability of trapped magnetic flux as discussed later. We also note that almost no anisotropic feature was observed in nonvolatile MTS for our Sn/Pb multilayers composites as discussed in the Supplemental Material [25] (see also references [26–28] therein). These results reveal that the formation of micro-scale domains of Sn and Pb is essential to induce nonvolatile MTS.

### C. Specific Heat and Magnetization Measurements

Figure 5(a) shows the results of specific heat measurement for the sample with $n$ = 13. After cooling to 2.0 K under zero magnetic field (ZFC) and under 1500 Oe (FC), the specific heat data were obtained at 0 Oe and with the inclement of 0.1 K from 2.0 K to 10.0 K. The plotted specific heat values for ZFC and FC in Fig. 5(a) represent the difference from that measured at 1500 Oe as the background signals for the normal conducting state. The green and red dashed lines at 3.7 K and 7.2 K correspond to the critical temperatures of Sn and Pb, respectively. In comparison of the results between ZFC and FC, a specific heat jump around 7.2 K is observed for both processes due to the superconducting-to-normal conducting transition of Pb. On the other hand, a specific heat jump around 3.7 K is observed only in the ZFC process, suggesting that the Sn domains remain in the magnetic-flux-trapping state throughout the entire temperature range for the FC process, thereby hindering the superconducting transition of Sn.

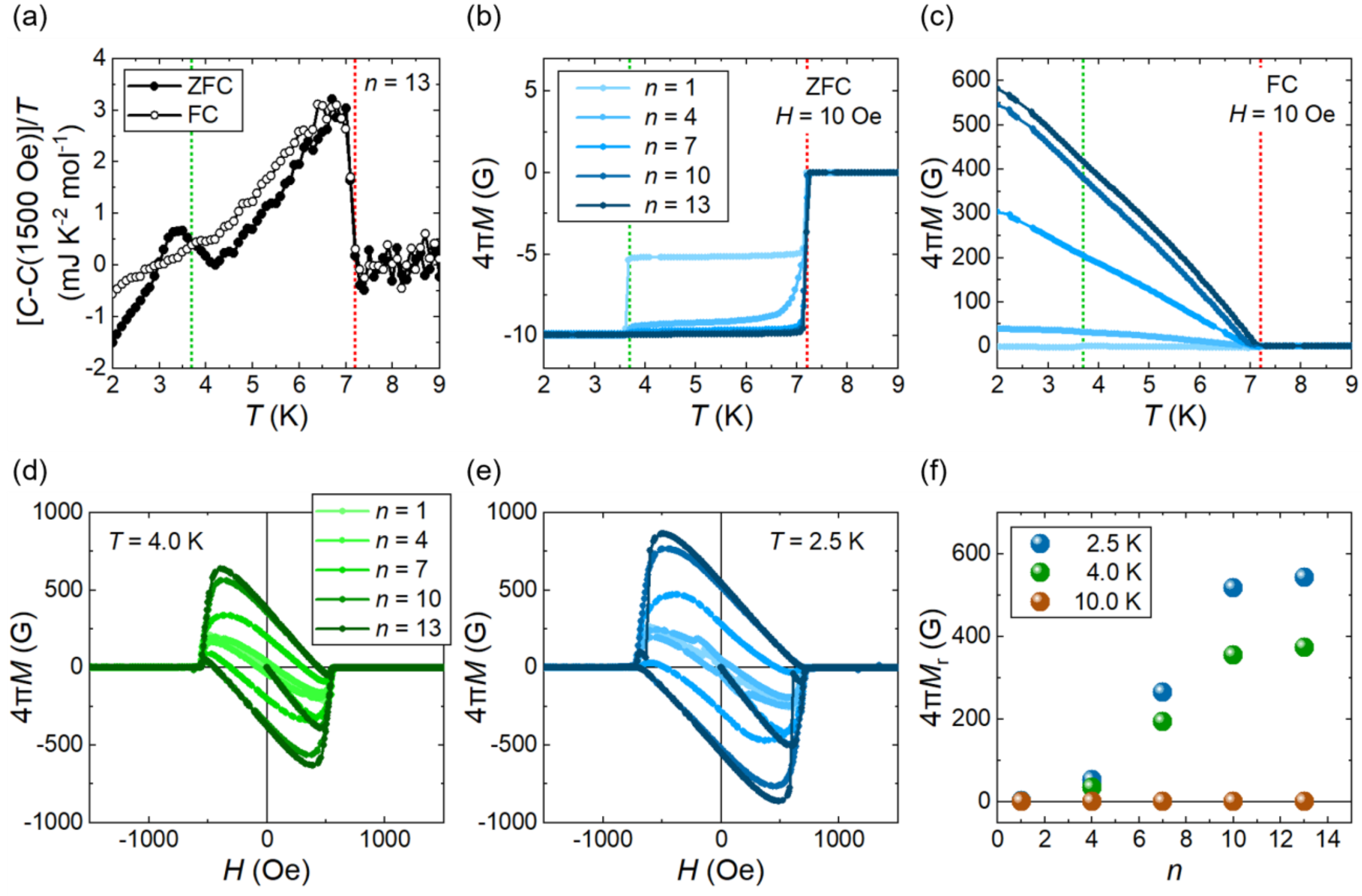

FIG 5. (a) Temperature $T$ dependence of the differential specific heat $[C\text{-}C(1500\text{Oe})]/T$ for the sample with the repetition number $n$ = 13 after the zero field cooling (ZFC) and field cooling (FC) processes. (b),(c) $T$ dependence of the magnetization $4\pi M$ after (b) ZFC and (c) FC processes for all the samples. $H$ during the measurements was set to +10 Oe. (d),(e) $H$ dependence of $4\pi M$ at (d) 4.0 K and (e) 2.5 K. (f) $n$ dependence of the remanent magnetization $4\pi M_r$ at 2.5, 4.0 and 10.0 K.

*Contact author: ANDO.Fuyuki@nims.go.jp, UCHIDA.Kenichi@nims.go.jp

Figures 5(b) and 5(c) show the temperature $T$ dependence of magnetization $4\pi M$ for all the Sn/Pb multilayered composites measured after the ZFC and FC processes. The magnetization measurements were performed after lowering $T$ to 2.0 K under zero magnetic field and 1500 Oe, setting $H = +10$ Oe, and sweeping $T$ to 10.0 K. $4\pi M$ for the ZFC process in Fig. 5(b) exhibits a perfect diamagnetism with $4\pi M$ of -10 G below both the critical temperatures of Sn and Pb for all the samples. Above 3.7 K, Sn undergoes a transition to the normal conducting state, causing a magnetization jump above -10 G. However, this phenomenon disappears as $n$ increases and the Sn and Pb region become finer because superconducting current is allowed to distribute throughout the entire sample and maintain $4\pi M$ of -10 G. On the contrary, for the FC process in Fig. 5(c), large positive magnetization is observed below the critical temperature of Pb, which derives from the trapped magnetic flux in the Sn region. $4\pi M$ at 2.0 K monotonically increases as $n$ increases and it decreases as $T$ increases due to the thermal instability of the trapped magnetic flux.

Figures 5(d)-(e) show the $H$ dependence of $4\pi M$ for all the Sn/Pb multilayered composites at 2.5 K and 4.0 K after the ZFC process. $H$ is swept with 4 Oe/sec from 0 Oe to +1500, -1500, +1500 and 0 Oe. As $H$ increases from the initial state, $4\pi M$ decreases due to the perfect diamagnetism and jumps to zero upon exceeding the critical field of Pb. Because both Sn and Pb are type-I superconductors, a perfect diamagnetism without hysteresis is expected. However, when $H$ is reduced back to zero, the samples with high $n$ exhibit the positive remanent magnetization $4\pi M_r$ with hysteresis rather similar to type-II superconductors. This is because microstructural refinement in the Sn/Pb multilayered composites enables the magnetic flux trapping in the Sn region. Figure 5(f) shows the $n$ dependence of $4\pi M_r$, which is originally zero at $n = 1$ but increases with increasing $n$ as well as nonvolatile MTSR in Fig. 4.

## IV. DISCUSSION

Let us discuss the possible reason why the Sn region can trap the magnetic flux below the critical temperature of Pb. Our experimental results on the microstructure dependence of nonvolatile MTSR fully supports the argument in ref. 19, which concluded that the size and proximity effects simultaneously contribute to the transition of Sn into a type-II superconducting state with an enhanced critical temperature associated with the magnetic flux trapping. We observed finite nonvolatile MTSR and $4\pi M_r$ even for the samples with $n = 4$ and 7, which respectively have average thicknesses of ~20 and ~2 μm. These values are larger than the critical thicknesses of Sn and Pb to transition into type-II superconductors [29,30] and the length of the proximity effect at junctions of superconductors and metals [31,32]. Thus, these mechanisms cannot solely explain our experimental results but synergistically contribute to them. Also, we note that magnetic flux trapping is essential for this transition because the Sn/Pb multilayered composites initially act as type-I superconductors after ZFC process regardless of the $n$ value [Fig. 5].

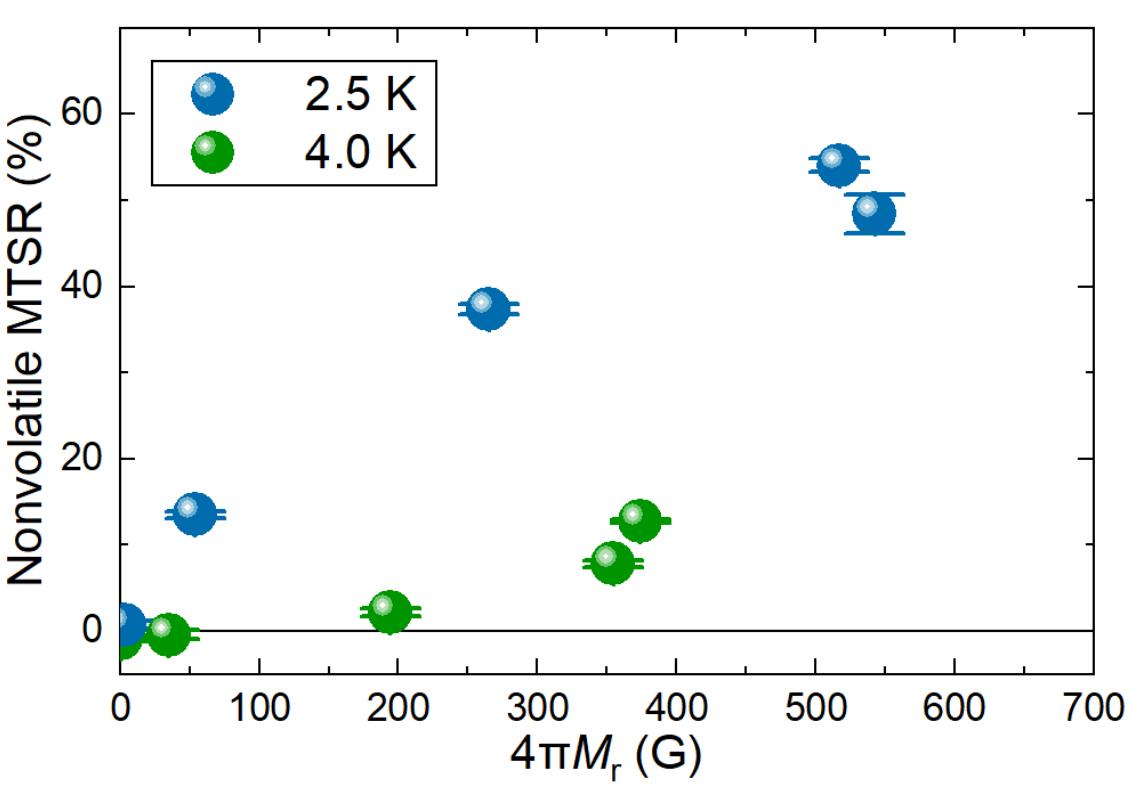

FIG 6. The relationship between the remanent magnetization $4\pi M_r$ and nonvolatile MTSR for the Sn/Pb multilayered composites at 2.5 and 4.0 K.

Figure 6 plots nonvolatile MTSR as a function of $4\pi M_r$, which clearly offers a positive relationship. Interestingly, even at the similar value of $4\pi M_r$, nonvolatile MTSR at 2.5 K is higher than that at 4.0 K. This difference implies that not only the absolute value of $4\pi M_r$ but also the spatial distribution of magnetic flux plays an important role on nonvolatile MTSR. The typical size of trapped magnetic vortex for the Sn-Pb system was reported to be ~10 μm at 5.0 K [19], which is comparable to the scale of microstructure in our Sn/Pb multilayered composites. The coherence length at 2.5 K is shorter than that at 4.0 K from the Ginzburg-Landau theory, which will enable the magnetic vortex to more homogeneously disperse in any size Sn domains and more efficiently destroy the neighboring superconducting domains of Pb. Thus, one may need to balance the microstructure size and coherence length of constituent superconductors to create the large nonvolatile MTS materials.

*Contact author: ANDO.Fuyuki@nims.go.jp, UCHIDA.Kenichi@nims.go.jp

## V. CONCLUSIONS

In this study, we demonstrate the nonvolatile MTS in superconducting Sn/Pb multilayered composites originating from the formation of the microscale Sn domains and their magnetic flux trapping. The multilayered Sn/Pb composite materials were systematically created with the controlled structural scale by the ARB process to study the microstructure dependence of MTS. We find that trapping of magnetic flux is enabled by the formation of domains whose size is comparable to or less than that of magnetic vortex. This work provides a distinct guideline to broaden the possibilities for developing new nonvolatile MTS materials utilizing superconductors for an efficient thermal management technology at cryogenic temperatures.

## ACKNOWLEDGMENTS

The authors thank Y. Hirayama and H. Arima for valuable discussion and H. Sebata, M. Maeda, and N. Kurata for technical support. This work was supported by ERATO "Magnetic Thermal Management Materials" (No. JPMJER2201) from Japan Science and Technology Agency, Iketani Science and Technology Foundation, by a Grant-in-Aid for Transformative Research Areas (A) "Correlation Design Science" (No. JP25H01248), and by a Grant-in-Aid for Scientific Research (KAKENHI) (No. JP26K21719) from Japan Society for the Promotion of Science.

*Contact author: ANDO.Fuyuki@nims.go.jp, UCHIDA.Kenichi@nims.go.jp

*Contact author: ANDO.Fuyuki@nims.go.jp, UCHIDA.Kenichi@nims.go.jp

Supplemental Material

## Revealing nonvolatile behaviors in magneto-thermal switching using microstructure-controlled superconducting composites

Keigo Ito,[1,2] Yui Sakamoto,[2] Hossein Sepehri-Amin,[2] Yuto Watanabe,[3] Poonam Rani,[3] Kumpei Imamura,[1,4] Takamasa Hirai,[1,2] Keisuke Hirata,[1,2] Shunsuke Mori,[1,2] Yusuke Nakanishi,[1,2] Kenichiro Hashimoto,[1,4] Takasada Shibauchi,[1] Yoshikazu Mizuguchi,[3] Ken-ichi Uchida,[1,2,*] and Fuyuki Ando[2,*]

[1]*Department of Advanced Materials Science, Graduate School of Frontier Sciences, The University of Tokyo, 5-1-5 Kashiwanoha, Kashiwa, Chiba 277-8561, Japan*

[2]*Research Center for Magnetic and Spintronic Materials, National Institute for Materials Science, 1-2-1 Sengen, Tsukuba, Ibaraki 305-0047, Japan*

[3]*Department of Physics, Tokyo Metropolitan University, 1-1 Minami-osawa, Hachioji, Tokyo 192-397, Japan*

[4]*Department of Physics, Kyoto University, Sakyo-ku, Kyoto*, Kyoto *606-8502, Japan*

*Contact author: ANDO.Fuyuki@nims.go.jp, UCHIDA.Kenichi@nims.go.jp

## Magnetic-field-angle dependence of nonvolatile MTSR

Our Sn/Pb composites have a possibility to show anisotropic thermal transport properties and nonvolatile magneto-thermal switching (MTS) owing to the multilayered structure. To investigate this, we measured the magnetic-field-angle $\theta$ dependence of the thermal conductivity $\kappa$ for a Sn-Pb multilayered composite. The Sn/Pb sample was created by the accumulative roll bonding (ARB) method as described in the main text with the repetition number $n$ of 6. Figure S1(a) shows a schematic of the $\kappa$ measurement setup [25–27]. To estimate $\kappa$, the steady-state heat flow method is adopted in the same manner as the experiments shown in the main text. Under various magnetic-field-angles $\theta$ ranging from 0º (in-plane) to 90º (out-of-plane) and 180º (reversed in-plane), the $H$ dependence of $\kappa$ was measured with increasing and decreasing $H$ at 0.6 and 2.5 K.

Figure S1(b) shows the result of the $H$ dependence of $\kappa$ at 2.5 K. The critical field is different depending on $\theta$; the critical field at the in-plane magnetic field configuration is higher than that at the out-of-plane one because the magnetic flux easily penetrates in the out-of-plane direction. However, the initial and final $\kappa$ values at $H = 0$ Oe are almost the same irrespective of $\theta$, suggesting that the amount of trapped magnetic flux is unchanged by the magnetization direction. Figure S1(c) plots nonvolatile MTSR as a function of $\theta$ measured at 0.6 and 2.5 K, obviously showing the absence of anisotropy in nonvolatile MTSR.

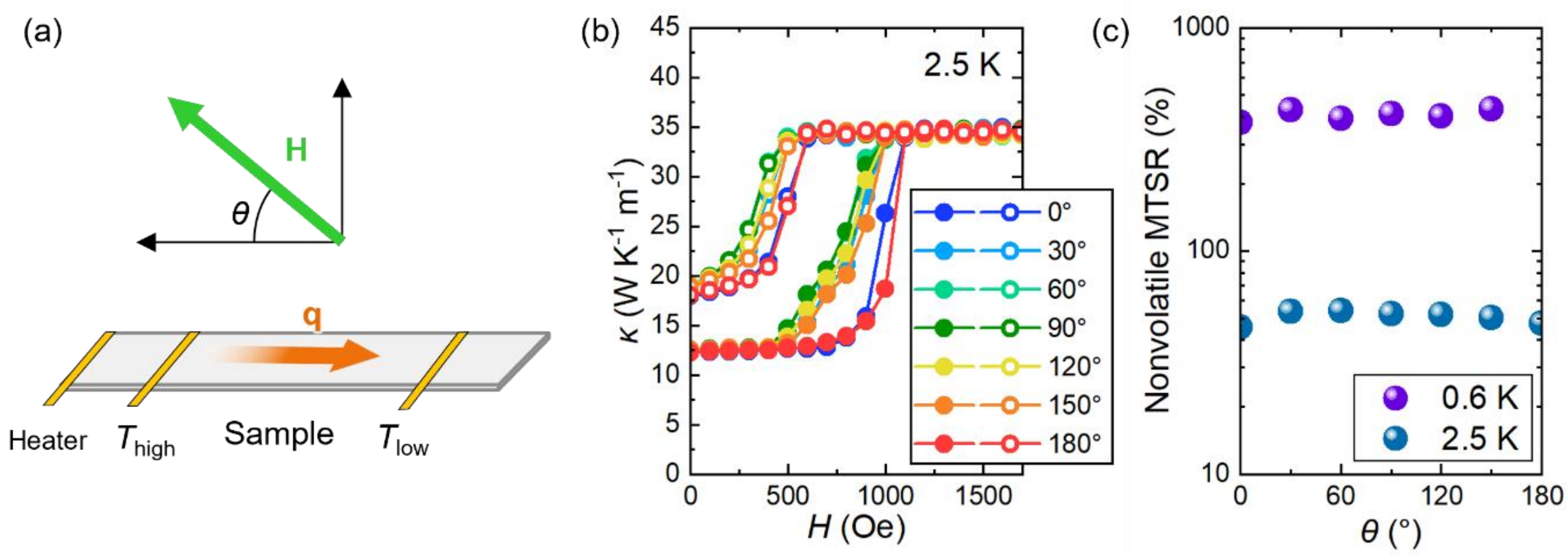

**Fig. S1** (a) Schematic of thermal conductivity $\kappa$ measurement setup based on the steady-state heat flow method, where heat flux **q** is applied to a sample under various magnetic-field ($\boldsymbol{H}$)-angles $\theta$ ranging from 0º (in-plane) to 90º (out-of-plane) and 180º (reversed in-plane). (b) $H$ dependence of $\kappa$ measured at 2.5 K. (c) Nonvolatile MTSR as a function of $\theta$ measured at 0.6 and 2.5 K.

*Contact author: ANDO.Fuyuki@nims.go.jp, UCHIDA.Kenichi@nims.go.jp